**Potential technosignature from anomalously low deuterium/hydrogen (D/H) in planetary water depleted by nuclear fusion technology**

David C. Catling[1,2], Joshua Krissansen-Totton[1,2], Tyler D. Robinson[2,3]

**Abstract**

Deuterium-deuterium (DD) fusion is viewed as an ideal energy source for humanity in the far future, given a vast seawater supply of D. Here, we consider long-lived, extraterrestrial, technological societies that develop DD fusion. If such a society persists over geologic timescales, oceanic deuterium would diminish. For an ocean mass and initial D/H that are Earth-like, fusion power use of only ~10 times that projected for humankind next century would deplete the deuterium-hydrogen ratio (D/H) in ~(a few) × $10^8$ years to values below that of the local Interstellar Medium (ISM). Ocean masses of a few percent Earth's would reach anomalously low D/H in ~$10^6$-$10^7$ years. The timescale shortens with greater energy consumption, smaller oceans, or lower initial D/H. Here, we suggest that anomalous D/H in planetary water below local ISM values of ~16×$10^{-6}$ (set by Big Bang nucleosynthesis plus deuterium loss onto dust or small admixtures of deuterium-poor stellar material) may be a technosignature. Unlike SETI from radio signals, anomalous D/H would persist for eons, even if civilizations perish or relocate. We discuss wavelengths of strong absorption features for detecting D/H anomalies in atmospheric water vapor. These are vibrational O-D stretching at 3.7 µm in transmission spectroscopy of Earth-like worlds, ~1.5 µm (in the wings of the 1.4 µm water band) in the shorter near-infrared for direct imaging by Habitable Worlds Observatory, and 3.7 µm or ~7.5 µm (in the wings of the broad 6.3 µm bending vibration of water) for concepts like the Large Interferometer for Exoplanets (LIFE).

(**250** words of 250 max)

[1]Department of Earth & Space Sciences, University of Washington, Seattle, WA 98195. USA
[2]Virtual Planetary Laboratory, University of Washington, Seattle, WA 98195. USA
[3]Lunar & Planetary Laboratory, University of Arizona, AZ 85721. USA

# 1. INTRODUCTION

The current telescopic search for life elsewhere focuses mainly on spectroscopic identification of potentially biogenic gases (e.g., $CH_4$ and $O_2$) emitted by microbes or plant-like organisms in extraterrestrial biospheres (Catling et al. 2018; Schwieterman et al. 2018), but technosignatures – detectable evidence of technology that modifies its environment (Tarter 2006)– may exist in a far wider range of planetary environments. While organic beings would likely originate on a conventionally habitable planet (Kitadai & Maruyama 2018), nonbiological robotic descendants of a technological society could tolerate broader environments and perhaps persist for far longer than their organic forebears (Shostak 2015). The inference of the ubiquity of habitable zone exoplanets (Bergsten et al. 2022), advances in detectors and signal processing, and the recognition of a wider range of possible environments for extraterrestrial technology, have sparked renewed interested in technosignatures, as reviewed by Wright et al.(2022).

Rather than the well-known searches for radio signals, some recent proposals of possible technosignatures consider influences on exoplanet atmospheric composition such as industrial pollutants. Specific gases proposed are nitrogen oxides from fossil fuels (Kopparapu et al. 2021) or chlorofluorocarbons (CFCs) (Haqq-Misra et al. 2022; Seager et al. 2023), but as trace gases they are generally undetectable with present or near-term technology. Also, such gases from immature civilizations would probably be extremely short-lived compared to the signal that we propose in this paper (Kopparapu, et al. 2021). For example, on Earth, after announcing CFC discovery in 1930 (Midgley Jr & Henne 1930), CFCs rose to concentrations that were a concern for ozone depletion by the 1970s, were phased out starting the late 1980s, and are now declining with atmospheric lifetimes $\sim 10\text{-}10^2$ years (Chipperfield et al. 2017). A centennial timescale of discovery to disappearance of such trace pollutants is one hundred millionth of the Sun's main sequence lifetime ($\sim 10^2/10^{10}$), and if such a timescale is similar for extraterrestrial civilizations, it implies a narrow temporal window for observing such technosignatures.

Nonetheless, the capabilities of telescopes to detect exoplanet atmospheric constituents should increase in the new few decades. Extremely large telescopes (ELTs) that are ground-based, such as the 39-m diameter European-ELT can use high dispersion spectroscopy to look at spectral features within narrow regions with high wavelength resolution (Snellen et al. 2015). Relevant spaceborne telescopes include the James Webb Space Telescope (JWST), launched in 2021, and developing mission concepts. The latter are NASA's Habitable Worlds Observatory (HWO), proposed for the 2040s in response to prioritization of direct imaging of Earth analogs (National Academies of Sciences & Medicine 2023), and the Large Interferometer For Exoplanets (LIFE) (Quanz et al. 2022). JWST can look at infrared spectral features of Earth analog exoplanet atmospheric gases in transit transmission spectroscopy if the planet is orbiting a low mass star (Kempton & Knutson 2024), whereas HWO is a coronagraphic space telescope to examine the spectrum of reflected light from exoplanets. The current HWO concept assumes sensitivity to ultraviolet, visible and near-infrared wavelengths below 2 μm. The LIFE concept currently proposes four 2-3.5 m telescopes as a nulling interferometer to obtain spectra of exoplanet emission within mid-infrared wavelengths (Glauser et al. 2024).

Here, we explore the consequences of nuclear fusion technology as a global energy source used by extraterrestrial societies over long timescales on the deuterium-to-hydrogen ratio in a major atmospheric gas, water vapor, on habitable worlds. Long-lived fusion technology would lower the D/H ratio of an ocean and atmospheric water sourced from evaporation could reach potentially anomalous levels. A D/H anomaly, if established, has the potential to last perhaps 7 or more orders of magnitude longer than the aforementioned technosignatures of pollution gases, perhaps shifting the probability of detection to the same extent.

It is reasonable to assume that global technological societies consume a lot of energy. In 2021, humankind used 595 EJ (exajoules, $10^{18}$ J) or an annually averaged power of 18.9 TW (BP 2022). Of this power, 15.6% (2.94 TW) was consumed in the US alone, such that the US population of 331.9M in 2021 had an energy-intensive lifestyle averaging 8.86 kW per capita continuously. To put such power into perspective, the sustained work capacity of an adult human male is ~75 W ( ~1/10 horsepower) (Webb 1973), so that each US person in 2021 had ~118 "human power equivalents" operating 24-7 across residential, industrial, agricultural, and transportation sectors to serve their needs.

From 1900 to 2000, the annual energy use of humankind increased ~9 times (Appendix in Smil (2010)), while by the end of the twenty-first century, the total energy demand will likely exceed present consumption by a factor of several, as follows. Figure 1 shows recent world annual energy use and a projection to ~2100 if the world's growing population trends towards US levels of per capita energy use. Here, we do not assume how this energy will be generated. Currently, humankind mostly uses fossil fuels for its primary energy and a only small proportion of renewables: 13% in 2021 (BP 2022). The fraction of energy use from fossil fuels must decrease to avoid climate disruption by global heating (IPCC 2023), but in any case, Figure 1 suggests that the annual power demand of the humankind could approach ~100 TW in the future if the world population grows as projected and US levels of per capita energy use become typical. Whether this level of power consumption happens near the end of the 21$^{st}$ century or far later depends on the uncertain rate at which per capita energy use will increase in developing countries.

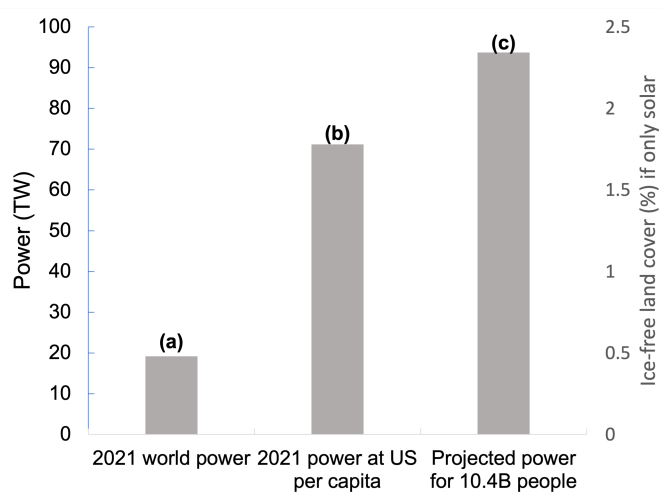

**Figure 1.** The annual average power use of humankind in Terawatts (left vertical axis) and the corresponding fraction of global, ice-free land area (right vertical axis) that would need to be covered by solar panels to provide all the power. We note that current human-modified land (infrastructure, crops, and managed land) leaves less than half of land in another state such as a desert or non-commercial forest (Hooke & Martin-Duque 2012). Also, a typical solar cell efficiency of 20% is assumed (Lewis 2007). **(a)** The total power in 2021 (BP 2022). **(b)** The power for the 7.9 billion human population of 2021 if everyone had used power at the same rate as Americans in 2021. **(c)** The additional power use for a population of ~10.4 billion assuming 2021 US per capita levels of power, which is the projected peak global population towards the end of the 21st century. The population projection is from the United Nations, Department of Economic and Social Affairs, Population Division: *World Population Prospects 2022, Online Edition*, https://population.un.org/wpp/ .

Even with 100% use of renewables or nuclear fusion in the long-term future of centuries or millennia, power consumption cannot increase indefinitely without planetary-scale climatic effects. For example, if the energy use of our descendants reached ~$10^4$ TW, the heat release and direct forcing of 19.6 W m$^{-2}$ would cause a very problematic ~15°C heating, using a typical equilibrium climate sensitivity of ~0.75 C/(W m$^{-2}$) (IPCC 2023). With an "Earth system sensitivity" >1 C/(W m$^{-2}$) that accounts for long-term climate feedbacks on timescales of millennia or more, such as shifting vegetation and ice sheet loss, the temperature increase would be even larger (Krissansen-Totton & Catling 2017; Ring, Mutz, & Ehlers 2022). Thus, power-hungry advanced civilizations may run into climatic limits on their power use regardless of the source, depending on the climate sensitivity of their planet. They would have to either curtail their energy use or deploy drastic geoengineering. In contrast, a global use of 1000 TW would release a global-average of ~2 W m$^{-2}$, comparable to the climate forcing that humans have already experienced because of $CO_2$ levels increasing by a factor of 1.5 from preindustrial 280 ppmv to ~420 ppmv in 2022 (IPCC 2023).

We hypothesize that the energy use of a global extraterrestrial society that is far more technologically advanced than humankind is likely to considerably exceed that of contemporary humankind because of multiple technologies, including artificial intelligence, automation or robotics, and technologies that we cannot yet imagine. We also assume that a more advanced civilization would solve the problem of nuclear fusion power because such technology would provide reliable energy over geologic timescales.

Indeed, nuclear fusion is arguably the most viable *in-situ* power source for long-lived, energy-intensive global societies. Renewable energy sources are theoretically sufficient for humankind's current and near-term power generation, but Earth-system limits preclude maintaining ~1000 TW production with wind, tidal, and geothermal sources (Miller et al. 2015). The maximum theoretical potential of *in situ* solar power is comparable to our assumed power demands (Kleidon, Miller, & Gans 2016) but would cause intolerable disruption to ecosystems from the huge land use. The size can be seen from scaling up an order of magnitude the land use in Figure 1, noting that over half of the land free of ice sheets is unavailable because of human-occupation or management. Nuclear fission is also not a viable long term power source because easily accessible uranium and thorium would be depleted (Abbott 2011). Similarly, long-term combustion of fossil fuels is unsustainable because of rapid fuel exhaustion and/or atmospheric oxygen depletion on geologic timescales, even on habitable planets for which the climate impacts are negligible (Graham 2021). It is possible to generate large amounts of power using off-world solar power—and technosignatures arising from transiting megastructures and Dyson swarms have been considered previously (Dyson 1966; Smith 2022; Wright 2020). But reliably and safely transferring ≥1000 TW to a planetary surface from space presents engineering challenges. In any case, we see no rational justification for an extraterrestrial society to stop using nuclear fusion once developed, given its exceptionally high energy density, small areal footprint, and reliability as a continuous source.

In nuclear fusion, the two promising reactions are the DT reaction, where deuterium ($^2$H or D) fuses with tritium ($^3$H or T) into helium-4, and the DD reaction, where deuterium fuses with deuterium to make helium-4 (reviewed by McCracken & Stott 2013). DT is currently being explored for contemporary fusion rather than DD because DT uses a lower ignition temperature to overcome Coulomb repulsion at attainable pressures and reach a threshold where self-heating from fusion just balances x-ray and bremsstrahlung radiative loss. For the DT reaction, the ignition temperature is 4.3 keV or ~5 x $10^7$ K whereas DD requires 35 keV or ~4 x $10^8$ K (Atzeni & Meyer-ter-Vehn 2004, p. 45; Post 1956). We ignore the idea of power generation via deuterium-producing proton-proton fusion because the collisional cross section is over 20 orders of magnitude smaller than DT or DD at similar fusion temperatures (Atzeni & Meyer-ter-Vehn 2004, p.12).

In the very long-term, on timescales discussed below, DD fusion is more attractive than DT because deuterium is available in seawater whereas tritium has a half-life of 12.33 years and must be generated from lithium. Tritium is made by bombarding a lithium blanket covering the inside of a fusion plasma chamber with energetic neutrons, which are produced after fusion gets going with some tritium seeding, making a tritium breeding cycle. Either $^6$Li or $^7$Li plus a neutron makes tritium and $^4$He, but $^6$Li has a far higher reaction cross-section than $^7$Li and is needed for sustained tritium breeding (Giegerich et al. 2019). Consequently, DT fusion requires $^6$Li and D as external fuels.

Supplies of lithium limit the lifespan of DT fusion power. Lithium ore currently totals 89 million tonnes worldwide (USGS 2022). In contrast, the ocean contains ~0.17 ppm Li (Bruland 1983) of which $^6$Li is 7.42% (James & Palmer 2000), i.e., a mass of (0.17×10$^{-6}$) ×0.0742× (1.426×10$^{21}$ kg/ocean) ~ 1.8×10$^{12}$ kg $^6$Li or ~18,000 million tonnes. Lithium could be extracted from seawater by electrolysis (Yang et al. 2018) followed by $^6$Li-enrichment. Each DT fusion reaction releases 2.8×10$^{-12}$ J. Because each atom of $^6$Li generates one tritium atom, the amount of energy per gram of $^6$Li is (2.8×10$^{-12}$ J × $N_A$)/( 6.015125 g/mol $^6$Li) = 2.8 ×10$^{11}$ J, where $N_A$ is Avogadro's number, 6.022×10$^{23}$. For an upper limit on longevity, we assume 100% energy conversion efficiency. With global power of 1000 TW, lithium-6 from ores would be used up in only ~60 yr (assuming 7.5% $^6$Li abundance in ore lithium (Faure & Mensing 2005, p. 854), while oceanic lithium-6 would last ~160 kyr. Of course, even if extraterrestrials found a way to supply the precursors of DT fusion for a longer period, e.g., from off-world mining, deuterium depletion of an ocean (as we calculate below) would still happen.

DD fusion is an inherently more long-lived source of energy by about three orders of magnitude for the same power consumption than DT fusion because of the oceanic supply of D, but such use of D over geologic timescales could deplete the D/H ratio of an ocean below natural D/H ratios (Figure 2). Big Bang nucleosynthesis (BBN) set the initial D/H ratio for the universe to an estimated (25.8±1.3)×10$^{-6}$ (Cyburt et al. 2016; Weinberg 2017). But stars destroy the deuterium they start with and so the D/H ratio in the Interstellar Medium (ISM) can be somewhat lower than the BBN value because of a contribution from a small fraction of D-depleted stellar material or because of hypothesized deuterium depletion on dust grains. However, models show that the D/H ratio in the ISM is unlikely to differ more than a factor of ~2 from the BBN value (Friedman et al. 2023). The measured D/H ratio of the local (<100 parsec) ISM is $(15.6\pm0.4)\times10^{-6}$ (Linsky et al. 2006), which sets a baseline minimum of local astronomical sources (Figure 2). If the D/H ratio in the water of an exoplanet was found to be substantially below ISM values – in the green shaded region of Figure 2 – it would be strange and anomalous.

In fact, the D/H ratio in the water of known rocky planets or icy moons is an order of magnitude or more than the protosolar D/H (in hydrogen) (Figure 2) because pre-stellar chemistry makes D-enriched water. In general, ices, such as those in comets or moons accreted from comet-like material (e.g., Enceladus), concentrate deuterium in both organic and water molecules because low-temperature reactions in pre-stellar clouds at temperatures of ~10 to ~100 K favor isotopic exchange reactions such as $H_2O$ + HD = HDO + $H_2$ (Geiss & Reeves 1981; Horner, Mousis, & Hersant 2007). Because rocky planets accrete their hydrogen as water ice or icy dust, the high initial D/H ratios observed on Earth relative to the ISM arguably should apply to water on rocky exoplanets too. Of course, giant planets will likely have D/H ratios similar to protosolar or ISM values (Marboeuf et al. 2018), whereas hot sub-Neptunes close to parent stars may have higher D/H in their hydrogen escape of isotopically light hydrogen over billions of years, e.g., factors of 1.2-1.7 (Gu & Chen 2023) or even $\sim10^2$ (Cherubim et al. 2024) are possible. But on rocky worlds, initial D/H in water may generally be about an order of magnitude higher than gas giant hydrogen and perhaps Earth-like because of inheritance from pre-stellar icy material.

Of course, some rocky planets may have even more elevated D/H in their water vapor than pre-stellar ices because of very substantial atmospheric escape compared to the Earth, as observed in the deuterium-enriched atmospheric water vapor on Mars or Venus (Figure 2). But the high D/H in these cases is not relevant as an initial value from which a lower D/H technosignature would be derived by nuclear fusion technology. As reviewed in Catling and Kasting (2017), Venus D/H evolved that way because of a runaway greenhouse effect to its current state and a surface temperature beyond the critical point of water, making life impossible its surface. Mars' atmosphere has been subject to all physical escape processes

(thermal, suprathermal, and impact erosion) because of its small size and mass (Lammer et al. 2020), rendering Mars dry, frozen, and possessing a tenuous atmosphere today, unsuitable for even simple life on its surface. Consequently, D/H that is higher than Earth's by factors of many or orders of magnitude probably indicates a planet that is problematic for habitability on geologic timescales.

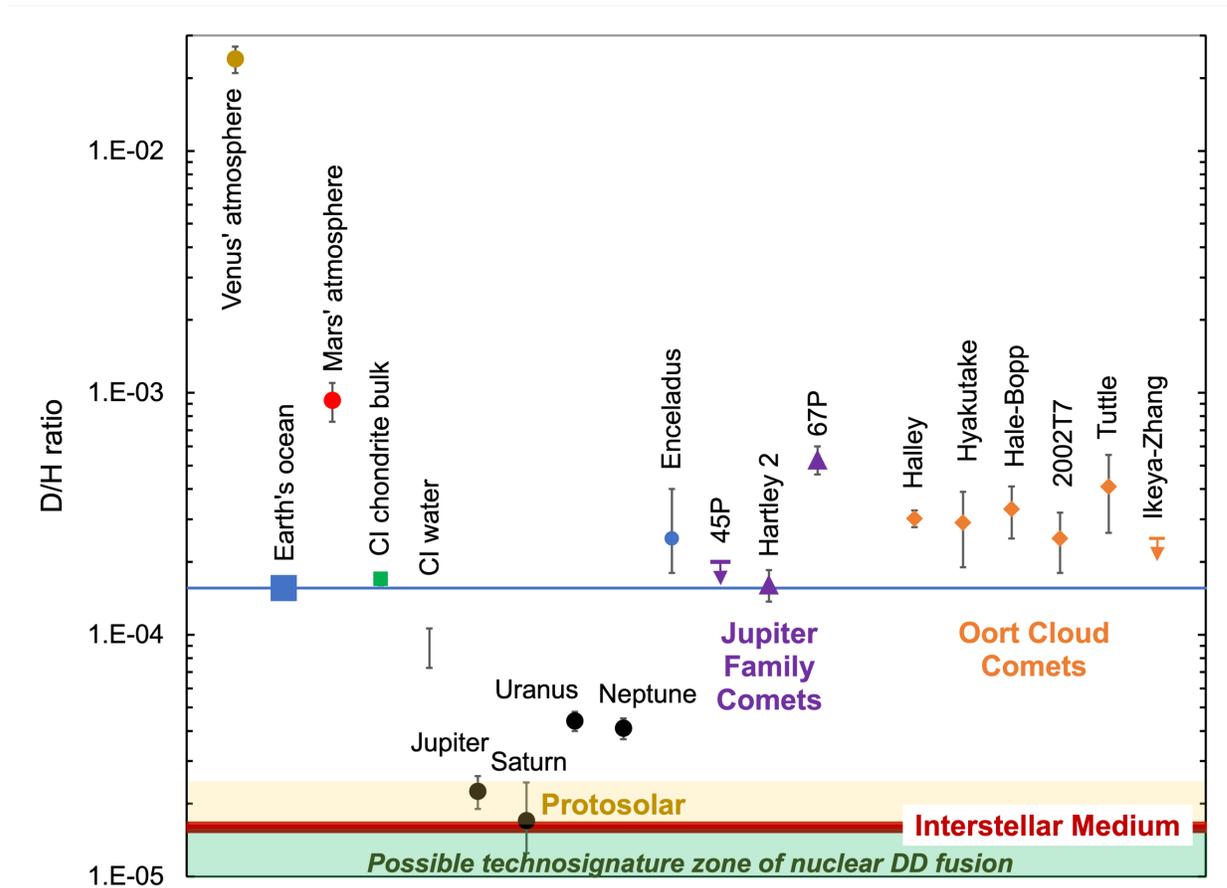

**Figure 2.** The deuterium/hydrogen (D/H) ratio in astronomical objects and on Earth. A technosignature in exoplanetary water of prolonged deuterium-deuterium nuclear fusion would plot below natural astronomical D/H values in the green shaded zone and be anomalous. Data sources: Bocklee-Morvan et al. (2015) and references therein for comets; Donahue et al. (1997) for Venus; Mahaffy et al. (2015) for Mars; Table S2 of Alexander et al.(2012)

for CI chondrites, and Lis et al.(2013) and references therein for other solar system objects, the interstellar medium, and protosolar D/H.

In the rest of the paper, we demonstrate proof-of-concept for the idea that the D/H ratio in planetary water could be depleted to anomalously low values below typical ISM values by extraterrestrial civilizations using DD fusion, even starting out with the equivalent of Earth's relatively high seawater D/H ratio. Depletion would be even more rapid if planets form with hydrogen inventories directly sourced from protoplanetary nebula hydrogen. We thus propose that future detection of unusually low D/H in exoplanetary water is a potential technosignature.

## 2. METHODS

2.1 Nuclear Fusion Assumptions

To calculate the depletion of deuterium in long-term nuclear fusion of an extraterrestrial civilization requires knowing an initial concentration of deuterium in an ocean and the energy release per gram of deuterium. For a practical reference calculation, we assume an initial D/H ratio of Earth's ocean. We then assume that a civilization uses fusion power over geologic time to calculate the resulting D depletion.

The D/H ratio of standard mean ocean water (SMOW) is 155.76±0.05 parts per million (Hagemann, Nief, & Roth 1970). So, a kilogram of water, which is 1000 g/(18.015 g/mol) = 55.51 moles of water = 2×55.51 moles of hydrogen, contains (155.76×10$^{-6}$)×2×55.51 = 0.01729 mol deuterium = 0.01729×(2.014 g/mol D) = 0.0348 g D per kg H$_2$O. Thus, there is ~35 g of deuterium per tonne (1000 kg) of seawater.

In DD nuclear fusion, D and D fuse into the following products with nearly equal probability ,

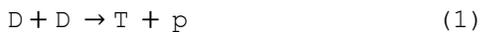
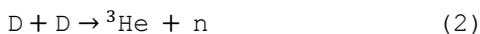

$$D + D \rightarrow T + p \qquad (1)$$
$$D + D \rightarrow {}^3He + n \qquad (2)$$

Then the products undergo subsequent fusion reactions, as follows,

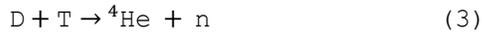
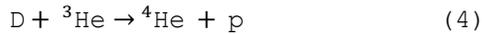

$$D + T \rightarrow {}^4He + n \qquad (3)$$
$$D + {}^3He \rightarrow {}^4He + p \qquad (4)$$

The overall net reaction in both cases, i.e., (1) + (3) or (2) + (4), is D + D + D → ⁴He + p + n. This net reaction produces slightly less energy compared to the idealized but unrealized reaction of D + D → ⁴He. The overall mass defect provides the energy release per gram of D. Working in Daltons (1 Da = 1.6605390666×10⁻²⁷ kg), the mass defect is $\Delta m = 3m_D - (m_{He} + m_p + m_n)$ = (3×2.0141) - (4.002603 + 1.007276 + 1.00866) = 0.023761 Da, where the $m$'s are CODATA-2018 masses with self-explanatory subscripts (Tiesinga et al. 2021). So, the energy = $\Delta m\ c^2$ = (0.023761× 1.6605390666×10⁻²⁷ kg) × (299792458 m s⁻¹)² = 3.5461×10⁻¹² J, where $c$ is the speed of light. Because this energy makes one ⁴He using three deuterium atoms, the energy per deuterium atom is one-third of this quantity, 1.182×10⁻¹² J. Hence, the energy per gram of deuterium is (1.182×10⁻¹² × $N_A$)/(2.01410 g D/mole) = 3.53 ×10¹¹ J per gram of D.

Any real nuclear fusion power plant will convert the energy released from fusion into usable energy with <100% efficiency, which some theoretical model estimate as ~30 to ~60% (Maisonnier et al. 2005). In our calculations, we assume 33% efficiency for conversion of fusion power into useful energy, so that the useful energy is ~1.178 ×10¹¹ J per gram of D. The net efficiency (i.e., electricity output relative to total energy entering) of coal and nuclear fission power plants has been ~1/3 for decades (US-EIA 2022), so our assumed efficiency is very conservatively grounded.

We also consider a range of average continuous power use for an extraterrestrial civilization. However, as described earlier, we adopt 1000 TW, or ten times the ~2100 estimate for power demand of humankind of ~100 TW (Figure 1), as a fiducial, conservative power use of extraterrestrial civilizations. This assumption might be considered extremely cautious compared to some oft-quoted, extravagant suggestions for the power use of advanced extraterrestrial civilizations. For example, ~10⁵ TW is commonly suggested for a Type I Kardashev civilization (Carrigan Jr 2010). However, Kardashev himself suggested that Type I was defined by

4 TW use for interstellar communications (Kardashev 1964), which was comparable to humankind's global power use of 4.6 TW in 1960 (Smil 2010). Sagan extended the definition of a Type I civilization to "using $10^{16}$ watts for interstellar communication" (Sagan 1973, p. 181). As noted previously, such a value of ~$10^4$ TW power for general energy demands would require very substantial geoengineering to ensure a livable climate system on an Earth-like planet.

The deuterium-hydrogen isotope system is the largest isotope difference for any chemical element and thus potentially remotely detectable in the HDO versus $H_2O$ isotopologues of water vapor in exoplanet atmospheres. Both ground-based and space-borne telescopes have been discussed for detecting the D/H ratio in exoplanet atmospheres (Lincowski, Lustig-Yaeger, & Meadows 2019; Mollière & Snellen 2019; Morley et al. 2019).

2.2 Radiative Transfer and Spectral Modeling

We explore the best wavelengths for detectability of deuterium depletion on atmospheric water vapor using a full-physics radiative transfer tool, the Spectral Mapping Atmospheric Radiative Transfer (SMART) model (Meadows & Crisp 1996; Robinson 2017). Preliminary simulations revealed that HDO depletion is best distinguished at 3.7 μm, which is the fundamental wavelength for the symmetric O-D stretching vibration ($v_1$ mode) in HDO. As this spectral region is challenging for both reflected- and emitted-light observations (due to low planetary photon count rates), our simulations initially explored transit spectroscopy for worlds with an atmospheric composition similar to a standard Earth atmosphere (McClatchey et al. 1971). High-resolution transit spectra were degraded to a resolving power ($\lambda/\Delta\lambda$) of 100, which is appropriate for the prism observing mode for the Near Infrared Spectrograph (NIRSpec) aboard JWST (Bagnasco et al. 2007).

With SMART, we also explored potential wavelengths for detecting deuterium depletion with the currently developing HWO and LIFE missions. The exact wavelength range for these missions is not finalized but HWO is likely to be sensitive to wavelengths that are smaller than 2 μm (to avoid instrument cooling) with a near-infrared resolving power ($\lambda/\Delta\lambda$) of 40-70. LIFE, as currently envisaged, has a nominal wavelength range of around 4 to 18.5 μm and resolving power 20–100 (Glauser, et al. 2024).

## 3. RESULTS

3.1 Feasibility of D/H depletion by fusion power

To put results into context and provide a check on calculations, we analytically estimated the expected longevity of DD fusion energy with terrestrial parameters and how long it would take to reach an anomalous D/H ratio that drops below natural astronomical D/H values. The total mass of deuterium in the ocean is (0.0348 g D/kg) ×(1.426×10$^{21}$ kg/ocean) = 4.96×10$^{19}$ g, and the total DD fusion energy available is (3.53 ×10$^{11}$ J/g) × (4.96×10$^{19}$ g) = 1.75×10$^{31}$ J (see Methods for derivation of the specific energy from DD fusion). For a sense of scale, this vast amount of energy is almost within an order of magnitude of the total gravitational binding energy of the Earth, ~2 × 10$^{32}$ J. Nonetheless, a long-lived civilization using DD fusion at 10$^3$ TW power with 1/3 efficiency and a terrestrial mass of ocean deuterium would exhaust all the deuterium for DD fusion after ~185 million years.

As an extraterrestrial civilization fuses oceanic deuterium into helium, the D/H ratio in the ocean would decline, while the helium would escape to space. We are interested in an anomalous D/H ratio below natural values, a threshold that we take as the local ISM D/H of 15.6×10$^{-6}$, which is ~10% of the initial D/H ratio in Earth's ocean (Figure 2). Thus, removal of ~90% of the D would reach this threshold, so we would expect the timescale to be ~0.9 × 185 Myr ~167 Myr for a 10$^3$ TW extraterrestrial civilization with an Earth-like ocean.

Thus, using the quantities discussed above in Methods, we can readily calculate the timescale required to reach a D/H ratio that matches local ISM value for a range of power usage to examine the plausibility of such a signal occurring. Figure 3 shows the sensitivity to ocean mass by considering cases of 0.5, 1.0, and 1.5 times the Earth's ocean mass. For our nominal 10$^3$ TW civilization, the timescale is ~167 million years, as expected, for one Earth ocean mass ($M_{ocean}$) or 1/20th, half, or 1.5 times this timescale for 0.05$M_{ocean}$, 0.5$M_{ocean}$ or 1.5$M_{ocean}$, respectively, from linear scaling. Thus, if we assume that an advanced extraterrestrial civilization and their descendants can persist for geologic timescales – which is likely a prerequisite for the success of many technosignature searches more broadly — then the generation of an anomalous D/H ratio from nuclear fusion is within the realm of possibility.

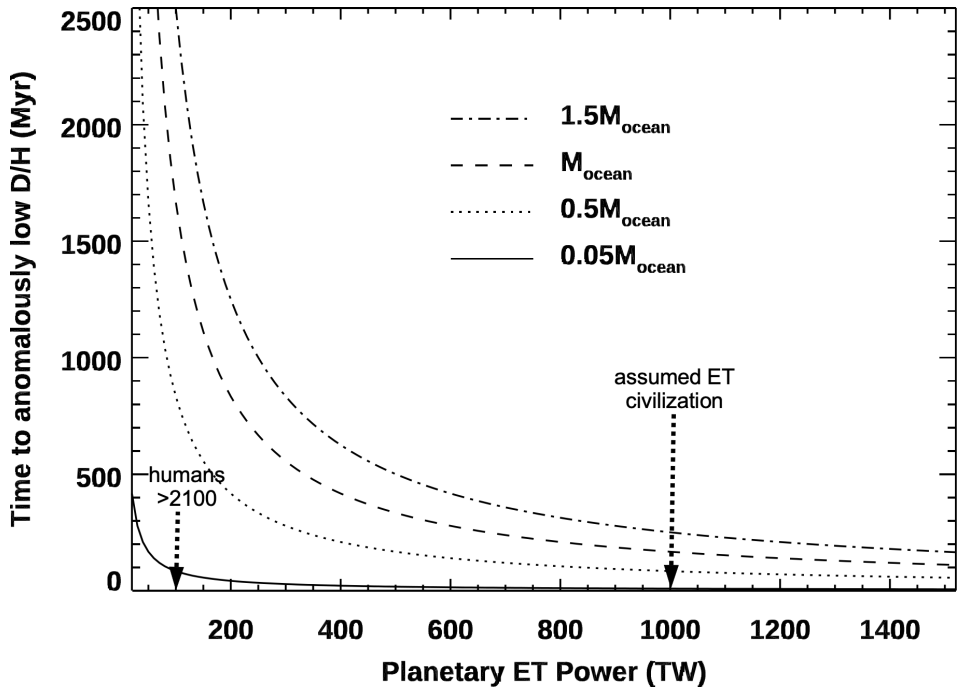

**Figure 3.** The timescale for a global extraterrestrial (ET) civilization using nuclear fusion power to reach a point beyond which D/H will become anomalously low compared to natural astronomical sources. Different ocean masses are assumed where $M_{ocean}$ is the ocean mass of Earth. Arrows indicate the possible projected power use of humankind beyond the end of the current century and an assumed ET civilization using 1000 TW continuously.

The timescale depends on the mass of the ocean, the energy consumption rate, and the initial deuterium content. The timescale becomes shorter the less massive the ocean, so that extraterrestrial nuclear fusion at $10^3$ TW on a so-called "land planet" with limited liquid water (Abe et al. 2011) would rapidly produce an anomalous D/H ratio, e.g., after only ~7 Myr for 4% of Earth's ocean mass. Even modest energy consumption (such as the that predicted for humankind with a population predicted for ~2100 and 2021 American per capita power use) would produce an anomalously low D/H ratio on a Gyr timescale that is less than the main sequence lifetime of a sun-like star. The initial D/H ratio of the ocean sets the mass of deuterium to be consumed, so a higher or lower initial D/H than the Earth's SMOW value of ~156 parts per million would cause the timescale to increase or decrease proportionately. Earlier, we presented a cosmochemical argument that sources of rocky planet water suggest an initial D/H that is may not be significantly dissimilar to Earth's for long-lived habitable worlds.

3.2 Platform-specific wavelengths to detect anomalous, fusion-depleted D/H

The potential shift of D/H to anomalously low values because of extraterrestrial nuclear fusion technology begs the question of whether it is detectable. Extremely large (40 m-class) ground-based telescopes could use high dispersion spectroscopy to measure the HDO isotopologue of several strong HDO lines near 3.6-3.7 μm in the reflected light of nearby rocky exoplanets, which is a wavelength that has been used to remotely detect D/H ratio in the atmosphere of Mars from ground-based telescopes (Krasnopolsky et al. 1997; Owen et al. 1988) and orbit (Vandaele et al. 2019). Molliere and Snellen (2019) calculate that 1 night of Extremely Large Telescope observing time with $\lambda/\Delta\lambda \sim$ 100,000 at ~3.7 μm would be sufficient to measure HDO on Proxima Cen b if it has an Earth-like atmosphere with Earth-like D/H ratio in its water vapor. Of course, a signal for >10 times depleted D/H would require longer integration.

Large spaceborne telescopes include JWST currently, the HWO concept proposed for the 2040s, and LIFE. Above Earth's atmosphere and with high resolution, many wavelengths have HDO lines, such as 2.63 μm, 3.7 μm and 8.1 μm, and there are also differences between $H_2O$ and HDO in ultraviolet absorption cross-sections (Cheng et al. 2004; Chung et al. 2001). Before JWST data indicated that the exoplanet TRAPPIST-1b likely has little or no atmosphere (Greene et al. 2023), JWST was shown to be suitable for searching for Venus-like D/H (~$10^2$ times Earth's) on TRAPPIST-1b using broad HDO bands at 3-4 μm, which produce a ~50 ppm signal in transmission spectroscopy that is potentially distinguishable from Earth-like D/H in ~11 transits at 5σ (Lincowski, et al. 2019).

Our radiative transfer calculations (see Methods) appropriate for the best JWST wavelength detection range is shown in Figure 4. This figure demonstrates the possibility of using transit spectroscopy to detect the effect of deuterium depletion in water vapor for an Earth-twin and a world with 10 times enhanced atmospheric water vapor, which could be the case for a warm Earth-like world near the habitable zone inner edge. Spectra emphasize the O-D stretching vibration of HDO near 3.7 μm. To simplify the example, models are cloud free and depleted cases have all deuterium-containing water vapor removed. Deuterium depletion reduces the effective transit altitude — which can be thought of as the altitude where unity slant-path optical depth is achieved — by 4-5 km in the O-D fundamental stretching vibration. For a late-type M dwarf with a radius of one-tenth a solar radius, the deuterium depletion impact on the transit spectrum translates to a transit depth difference of roughly 10 ppm, which is of a scale that could be detected with an observatory like JWST (Rustamkulov et al. 2022). To lower noise to a level where the D/H feature at 3.7 micron might be discerned, in principle, at resolving power of 100, would take an integration of 43 hours, but considering additional outside-of-transit time, it is more realistically ~100 hour or ~40 transits.

Future models for transmission spectroscopy and the other observing platforms discussed below, beyond the scope of this proof-of-concept paper, could consider influences of clouds, three-dimensional atmospheric circulation and structure, star spots, instrument noise models, and variable deuterium depletions for more detailed estimates of observability.

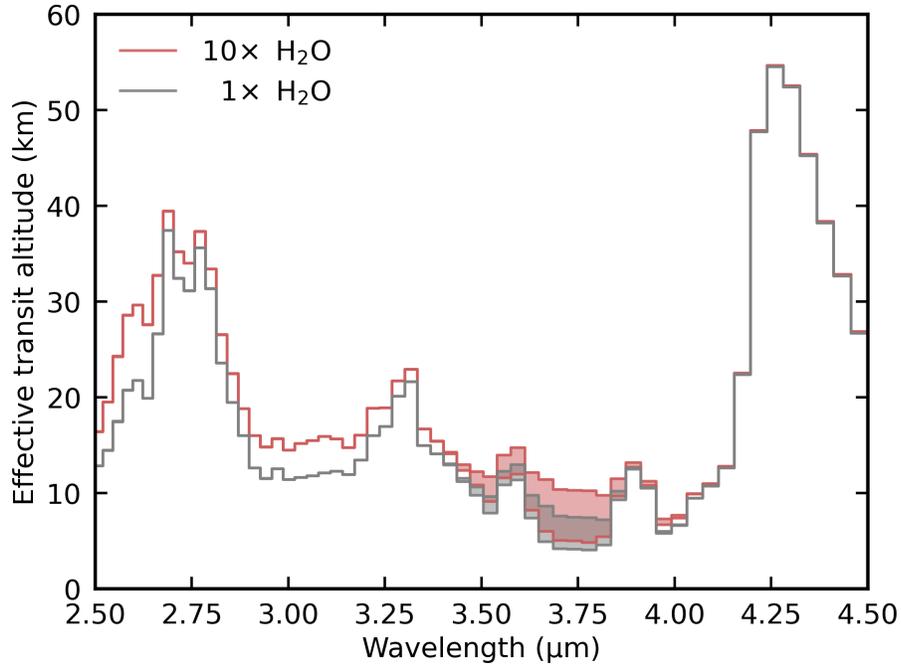

**Figure 4**. Transit spectra of deuterium-depleted Earth-like worlds, showing a potentially detectable effect at the fundamental vibrational stretch mode for O-D in HDO near 3.7 μm. Cases considered are an Earth-twin atmosphere (gray) and an atmosphere where water vapor has been enhanced at 10 times the Earth-twin value (red). Filled swaths show the impact of removing all deuterium-containing water vapor from the atmosphere in each case.

HWO is still being defined but will involve direct imaging of exoplanets and reflection spectroscopy, though revealing HDO features at longer near-infrared wavelengths would require cooling many elements of the telescope system. A search for the best wavelength to measure HDO in water vapor in reflected light at HWO wavelengths (i.e., smaller than 2 μm) revealed a feature around 1.5 μm, in the wing of the 1.4 μm water vapor band, shown in Figure 5(a). The corresponding signal-to-noise ratio (SNR) for the case of high water vapor abundance is shown in Figure 5(b) for one sigma detection.

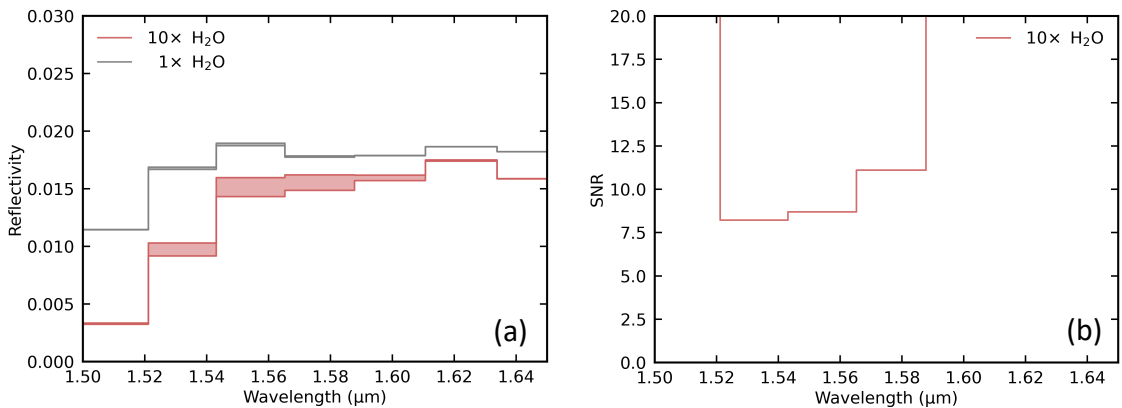

Figure 5. (a) Reflectivity of deuterium-depleted Earth-like worlds, showing a features from HDO near 1.5-1.6 µm. Cases considered are an Earth-twin atmosphere (gray) and an atmosphere where water vapor has been enhanced at 10 times the Earth-twin value (red). Filled swaths show the impact of removing all deuterium-containing water vapor from the atmosphere in each case. (b) The signal-to-noise (SNR) required for one sigma detection. An HWO-like resolving power of 70 is adopted.

A similar search of wavelengths in the mid-infrared, relevant to the Large Interferometer For Exoplanets (LIFE) concept, showed that ~3.7 µm and a ~7.5 µm feature in the wings of the broad bending vibration band at 6.3 µm could detect HDO depletion at SNRs < 10. Figure 6a shows the full infrared wavelength range modeled, whereas Figure 6b illustrates how 3.7 µm requires a lower SNR than ~7.5 µm for detection deuterium depletion.

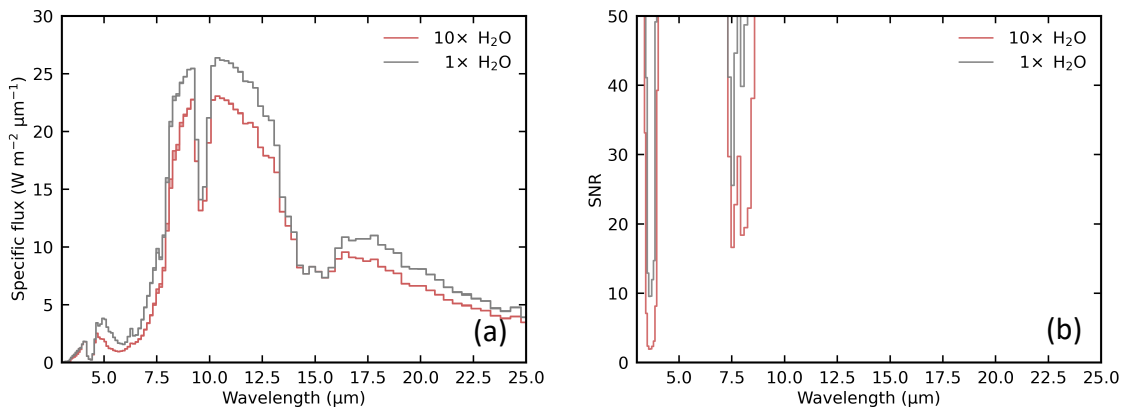

Figure 6. (a) Features of deuterium-depleted Earth-like worlds across the mid-infrared. Again, cases considered are an Earth-twin atmosphere (gray) and an atmosphere where water vapor has been enhanced at 10 times the Earth-twin value (red). Filled swaths show the impact of removing all deuterium-containing water vapor from the atmosphere in each case. (b) The signal-to-noise (SNR) required for one sigma detection. A LIFE-like resolving power of 50 is adopted.

## 4. DISCUSSION

We have demonstrated that anomalous D/H depletion by extraterrestrial nuclear fusion technology is conceivable (Figure 3) and would generate features at near- and mid-infrared wavelengths that should be considered for current (Figure 4) and future telescopes (Figures 5 and 6). In future measurements of D/H, it may be important to consider possible climatic influences on the D/H in atmospheric water vapor, specifically the preferential rainout of relatively isotopically heavy ice, rain or snow that might lower D/H in water vapor. Certainly, D/H in atmospheric water vapor on planets cool enough to have icy poles would be D-depleted by condensation in polar regions by Rayleigh distillation extraction of D into condensates. Because of such effects, the disk average of the Earth's tropospheric water vapor is perhaps ~85-90% of the D/H in SMOW (Good et al. 2015; Worden et al. 2019; Zakharov et al. 2004). This ratio remains roughly an order of magnitude higher than the local ISM value, because the ISM value is ten times smaller than SMOW (Figure 2), but future models of hypothetical exoplanetary hydrological cycles with hydrogen isotopes might reveal the extent to which spatiotemporal patterns in isotopic ratios could modify disk averages. For example, a view looking directly down on an icy pole of a planet would skew the atmospheric D/H ratio to low values because of the temperature dependence of the fractionation and progressive rainout of heavy isotopologues in polar regions (Dansgaard 1964). Similarly, whether gradual dilution of surface D/H depletion from mantle degassing is possible is a subject for future modeling.

The demand that a fusion-induced D/H anomaly be below that of the ISM value may be too stringent. If the D/H ratios of several watery rocky exoplanets were measured in the future and found to be of the same order of magnitude as the Earth's ocean, a single D/H of water on a rocky world between that of Earth and the ISM might be considered anomalous. It might then be reasonable to consider whether such a D/H was a possible fusion technosignature. In this case, a sub-ISM value might not be needed, although it always would more definitive.

Compared to other technosignatures, a strong temporal advantage of using anomalously low D/H in rocky planet atmospheric water vapor is that the signal should persist for eons regardless of whether an extraterrestrial civilization has perished, relocated, or transitioned to another form of energy production. This extreme longevity makes our proposed technosignature qualitatively different from some other technosignatures. Radio or optical communications require active, extant extraterrestrial intelligence, but an anomalously low D/H ratio does not. Moreover, the non-detection of anomalously low D/H on a sufficiently large sample of potentially habitable planets could constrain the distribution of energy-consuming civilizations across time and space.

## 5. CONCLUSIONS

In this paper, we have presented a new idea that very long-lived, advanced extraterrestrial (ET) civilizations using continuous deuterium-deuterium (DD) nuclear fusion could lower the D/H ratio in an ocean to less than the ratio found in natural astronomical sources. Big Bang nucleosynthesis set an initial D/H ratio of 24-27 ppm but the average D/H in the local interstellar medium (ISM) of ~16 ppm is lower because of a small amount of mixing with deuterium-poor stellar material and hypothesized reactions on dust grains that take up deuterium. So, D/H values in water on exoplanets below the ISM baseline of natural astronomical sources would be anomalous.

Indeed, we expect the initial water on rocky planets to be D-enriched relative to the ISM because reactions in pre-stellar molecular clouds tend to concentrate deuterium in icy material and dust that contributes to the water accreted when rocky planets form. The known D/H ratios of water on rocky worlds in the solar system – Earth, Venus, and Mars – are far higher than the protosolar or ISM D/H ratio because the planetary water started at a higher ratio and all D/H values have probably increased due to atmospheric escape, particularly on Mars and Venus (Donahue, et al. 1997; Mahaffy, et al. 2015; Zahnle, Gacesa, & Catling 2019).

The timescale for DD fusion by an ET civilization to reach a D/H as low as the ISM – a point beyond which D/H would become anomalously low – depends on the power demands of the civilization, the mass of the ocean on their planet, and the initial D/H in the water. With an Earth-like initial D/H ratio and ocean mass, the timescale to anomalous depletion would be ~170 Myr for a civilization operating at $10^3$ TW. Ocean masses of a few percent Earth's would be anomalously depleted in $<10^7$ years at $10^3$ TW power.

The D/H ratio is potentially remotely detectable in the atmospheric water vapor of exoplanets with many strong HDO lines in the infrared and near-infrared. The fundamental O-D stretching vibration at 3.7 μm is the strongest feature accessible to JSWT. Furthermore, an explicit goal to measure D/H might be a justification to push for a slightly smaller cut-off wavelength than 4 μm for an interferometry mission such as LIFE. In reflected light below 2 μm, applicable to HWO, a feature around 1.5-1.6 μm is the best wavelength but requires challengingly high SNR. However, such a D/H signature should persist irrespective of whether an extraterrestrial civilization has ceased or left its planet. Finally, if a D/H technosignature is not found, measuring D/H of water on habitable zone exoplanets would provide information about sources and evolution of water, as D/H does in solar system studies (McCubbin & Barnes 2019).

**Acknowledgments**

This paper arose from a call from Jason Wright for "SETI-related ideas" to present at the *Blumberg Astrobiology Workshop* 2023, sponsored by Unither Bioelectronique, May 5-7, 2023, at the Green Bank Observatory. We thank Martine Rothblatt for her support of the workshop, which resulted in our thoughts and research in this paper. DCC, JKT and TDR were supported by the Virtual Planetary Laboratory under NASA grant 80NSSC18K0829 as part of the Nexus for Exoplanet System Science (NExSS) research coordination network.